\documentclass[12pt]{article}

\usepackage{graphicx}

\begin{document}

\title{Approach to Vibrational to Axially Rotational Shape Phase Transition
Along the Yrast Line}

\author{{Yu-xin Liu$^{1,2,3}$, Liang-zhu Mu$^{1}$}\\[3mm]
\normalsize{$^1$ Department of Physics, Peking University, Beijing 100871, China}\\
\normalsize{$^2$ Institute of Theoretical Physics, Academia
Sinica, Beijing 100080, China} \\
\normalsize{$^3$ Center of Theoretical Nuclear Physics, National
Laboratory of}\\ \normalsize{ Heavy Ion Accelerator, Lanzhou
730000, China} }


\maketitle

\begin{abstract}
With the energy surface of the nucleus in U(5) symmetry being
analyzed in the framework of thermodynamics, the vibration and
rotation phase diagram in terms of the angular momentum and
deformation parameter is given. Together with examining the energy
spectrum, we propose a theoretical approach to describe the
vibrational to axially rotational phase transition along the yrast
line. By analyzing the available experimental data we show that
the vibrational to rotational shape phase transition along the
yrast line takes place in many nuclei.
\end{abstract}

\bigskip




\newpage

\parindent=20pt

It has been well known that shape phase transition is one of the
most significant topics in nuclear structure research. Many
evidences of nuclear shape phase transition have been observed.
For instance, in several isotopes, there exists shape phase
transition from vibration to axial rotation or $\gamma$-unstable
rotation with respect to the variation of neutron
number\cite{IA87}, and a triple point may
appear\cite{Jolie02,Warn02}. Even in one mode of collective motion
there may involve different characteristics, for example, along
the yrast line there exists transition between rotations holding
different relations between angular momentum and rotational
frequency (referred to as band crossing and exhibiting a
backbending)\cite{SS72}. Very recently it was found that there
involves evolution from vibration to rotation along the yrast line
of a nucleus\cite{Regan03}. On the theoretical side, the
interacting boson model (IBM) has been shown to be successful in
describing the shape phase transition along a series of
isotopes\cite{IA87,Cejn03}. And analytic solutions for the
critical points of the phase transitions have been
found\cite{Iac00,Iac01,LG03}. The cranked shell model
(CSM)\cite{BRM86} has been known to be able to describe the band
crossing very well. However, a theoretical approach to describe
the shape phase transition from vibration to rotation along the
yrast line in a nucleus has not yet been established. Especially
the phase diagram in terms of the angular momentum and the
deformation parameter has not yet been obtained. Even so, the
energy level statistics analysis has shown that the states in the
U(5) symmetry with $ A < 0$ involve the coexistence of vibrational
and axially rotational shapes\cite{SRJL03}. By analyzing the
energy surface and the energy spectrum of the U(5) symmetry, we
will show that the U(5) symmetry with a special choice of
parameters can be an approach to describe the shape phase
transition along the yrast line and such kind phase transition
takes place in many nuclei.

In the original version of the IBM (IBM1), the collective states
of nuclei are described by the coherent states of $s$- and
$d$-bosons. The corresponding dynamical group is U(6), and
involves three dynamical symmetry limits U(5), O(6) and SU(3).
Taking into account one- and two-body interactions among the
bosons, one has the Hamiltonian of the nucleus in U(5) symmetry
as\cite{IA87}
$$\displaylines{\hspace*{1cm}
H_{U(5)}=E_0 + \varepsilon_{d}  C_{1U(5)} + A C_{2U(5)} + B
C_{2O(5)}+ C C_{2O(3)} \, , \hfill{(1)} \cr }
$$
where $C_{kG}$ is the $k$-rank Casimir operator of the group $G$
and the parameters hold relation $C \ll | B | \ll \vert A \vert
\ll \varepsilon _d $. Making use of the coherent state
formalism\cite{GK80,DSI80,IC81} of the IBM, one can express the
energy functional for the U(5) Hamiltonian as
$$\displaylines{\hspace*{1cm}
E(N, \varepsilon, A; \beta) = E_{0} + \varepsilon N
\frac{\beta^{2}}{1+\beta^{2}} + A
N(N-1)\frac{\beta^{4}}{{(1+\beta^2)}^{2}} \, , \hfill{(2)} \cr }
$$
where $\beta$ is a monotonous function of the usual deformation
parameter (for example, for the rare earth nuclei, the usual
quadrupole deformation parameter $\beta _2$ and the $\beta$ hold
relation $\beta _2 \approx 0.16 \beta$\cite{IA87}) and the
$\varepsilon$ can be expressed in terms of the parameters in
Eq.~(1) as $\varepsilon = \varepsilon _{d} + 5 A + 4B + 6C$. If
$\beta ^2$ is small (less than one), Eq.~(2) can be rewritten as
$$\displaylines{\hspace*{3mm}
E(N, \varepsilon, A; \beta) = E_{0} + \varepsilon N \beta^{2} + [
A N (N-1) - \varepsilon N] \beta^{4} + [\varepsilon N - 2 A N(N-1)
] \beta^{6} + \cdots \cdots  \, , \hfill{(3)} \cr }
$$
On the basis that the ``deformation parameter" $\beta$ is a
quantity to characterize the shape of a nucleus, one can take the
$\beta$ as the order parameter to identify the nuclear shape phase
transition.

In the general framework of Landau's phase transition
theory\cite{LL01}, it has been pointed out that, if the Landau
free energy $F(T, \xi)$ can be expressed as
$$\displaylines{\hspace*{2cm}  F(T, \xi) = F_{0} + \frac{1}{2}
\alpha (T- T_0) \xi ^2 + \frac{1}{4} g_4 \xi ^4 + \frac{1}{6} g_6
\xi ^6 + \cdots \cdots \, \hfill{(4)} \cr } $$
 where $T$ is the control quantity, $\xi$ is the order parameter,
$\alpha$ is a constant. It has been known that, if the other
coefficients which depend also on the control quantity $T$ hold
relation $g_4 < 0$, $g_6
> 0$, the system may involve a first order phase
transition\cite{KKbook}.

For a nucleus involving vibration to rotation phase transition
along the yrast line, the control quantity is the angular momentum
$L$. Comparing Eq.~(3) with Eq.~(4), we propose that the Eq.~(3)
can be rewritten  as
$$\displaylines{\hspace*{3mm}
E(N, \varepsilon, A; \beta) = E_{0} + \frac{1}{2} \alpha (L - L_0)
\beta^{2} - [  \frac{1}{2} \alpha (L - L_0) - A N(N-1) ] \beta^{4}
\hfill{} \cr \hspace*{4cm} + [\frac{1}{2} \alpha (L - L_0) - 2 A
N(N-1) ] \beta^{6} + \cdots \cdots \, , \hfill{(5)} \cr }
$$
where $AN(N-1)$ relies also on the control quantity $L$. It is
evident that, if $\alpha < 0$ (i.e., $\alpha(L-L_0) > 0$ for
$L<L_0$) and $A<0$, Eq.~(5) stands for the free energy of a
nucleus just the same as Eq.~(4) with $g_4 < 0$, $g_6 > 0$. Even
if $A>0$, we may have $\frac{1}{2} \alpha (L - L_0) - 2 A N(N-1) >
0$, Eq.~(5) corresponds also to Eq.~(4) with $g_4 <0$ and $g_6
>0$. Analyzing the free energy in Eq.~(5) more cautiously, one can
recognize that there exists an angular momentum $L_0$ with which
the $\frac{\partial ^2 E(N, \varepsilon, A; \beta)}{\partial \beta
^2}=0 $.  And $\alpha (L-L_0) > 0$, $\frac{\partial ^2 E(N,
\varepsilon, A; \beta)}{\partial \beta ^2} > 0 $, if $L < L_0$;
$\alpha (L-L_0) < 0$, $\frac{\partial ^2 E(N, \varepsilon, A;
\beta)}{\partial \beta ^2} < 0 $, if $L > L_0$. Meanwhile, there
exists an angular momentum $L_c = L_0 - (\frac{4}{3}\sqrt{3} -2)
\frac{AN(N-1)}{\alpha}$ (if $A<0$) with which the free energy
involves one maximum and two equal minima, one of which is
generated at $\beta =0$, another corresponds to $\beta \ne 0$.
There exists also a maximal angular momentum $L_{max} = L_0 +
(\sqrt{6} + 2)\frac{AN(N-1)}{\alpha}$ (if $A<0$). If $L\ge
L_{max}$, the energy surface has only one maximum at $\beta = 0$
but no minimum. The $\beta$ dependence of the free energy at some
typical angular momenta is illustrated in Fig.~1. The figure shows
apparently that, as the angular momentum $L < L_c$ the nucleus has
a stable vibrational phase (the free energy takes the smaller
minimum at $\beta =0$). If the angular momentum $L_c \le L < L_0$,
the nucleus has a stable rotational phase (the free energy takes
the smaller minimum at $\beta \ne 0$). Meanwhile there may involve
coexistence of rotation and vibration, since the energy surface
holds two minima at $\beta =0$, $\beta \ne 0$, respectively. If
the angular momentum $L_0 < L <L_{max}$, the nucleus has only a
rotational phase. The angular momentum $L_c$ is definitely the
critical angular momentum for the phase transition from vibration
to rotation to happen. In addition, if $L\ge L_{max}$, the nucleus
will collapse. We obtain then explicitly how the stable phase of a
nucleus is determined by the angular momentum and the
``deformation parameter". The phase diagram in terms of the
angular momentum $L$ and the ``deformation parameter" $\beta$ can
thus be displayed in Fig.~2. We reach thus a conclusion that the
U(5) symmetry with $A < 0$ can describe well the vibrational to
rotational nuclear shape phase transition along the yrast line.
And the transition is a first order phase transition. Even if
$A>0$, the first order phase transition may also happen in the
system if the interaction strengths satisfy the constraint
mentioned above. It manifests that, the vibrational to rotational
phase transition along the yrast line can take place in the
nucleus holding the U(5) symmetry in IBM1.

On the phenomenological side, the energy spectrum of a nucleus in
U(5) symmetry can be given as
$$\displaylines{\hspace*{1cm}
E_{U(5)}=E_0 + \varepsilon _{d} n_d + A \, n_d(n_{d}+4) + B\,
\tau(\tau+3)+ C \, L(L+1) \, . \hfill{(6)} \cr }
$$
where $n_d$, $\tau$, and $L$ are the irreducible representations
(IRREPs) of the group U(5), O(5) and O(3), respectively.

From the spectrum generating process one can recognize that, if
$A>0$ and $B>0$, the states with $\tau = n_{d}$, $L = 2 n_{d}$
form the ground state band and the yrast band, and appear as the
anharmonic vibrational ones with increasing frequency
$\hbar\omega=\varepsilon _{d} + (n_{d}+4) A $. If $A<0$, the
$E_{gsb}(n_d)$ ($E_{gsb}(L)$) is an upper-convex parabola against
the $n_d$ ($L$), and can describe the collective backbending of
high spin states\cite{Long97}. In the case with $A<0$, $B<0$, the
freedom of $\tau$ in Eq.~(6) can be washed out. It turns out then
the state with $\tau = n_{d}$ may still be the yrast one and there
exists a d-boson number $n^{(c)}_{d}$ and an angular momentum
$$\displaylines{\hspace*{1cm}
L_{c}=2n_{d}^{(c)} = - \frac{2(\varepsilon_{d} + 4 A + 3 B )}{A +
B } - 2N_{0} \, , \hfill{(7)} \cr }
$$
where $N_0=N$ with $N$ being the total boson number. As the
angular momentum $L \geq L_c$, the yrast states are no longer the
anharmonic vibrational ones mentioned above, but the
quasi-rotational ones with $\tau = n_{d}= N_0$. It indicates that
the U(5) symmetry with parameters $A<0$ and $B<0$ may describe the
vibrational to rotational phase transition along the yrast line.
The $L_c$ given in Eq.~(7) is the critical angular momentum. For
each yrast state with angular momentum $L \le L_{c}$, its energy
can be given as
$$\displaylines{\hspace*{1cm}
E(L \le L_c)=E_0 + \frac{ A + B + 4 C}{4} {L}^{2} +
\frac{\varepsilon_{d} + 4 A + 3 B + 2 C} {2} L \, . \hfill{(8)} \,
\cr }
$$
and can be illustrated as a part of an upper-convex parabola.
Whereas for the one with angular momentum $L > L_{c}$, its energy
should be expressed as
$$\displaylines{\hspace*{1cm}
E(L> L_{c} ) = E_{0} ^{\prime } + C L ( L +1)  \, , \hfill{(9)}
\cr }
$$
and can be displayed as a part of an upper-concave parabola. For
instance, for a system with $N=15$ and parameters $\varepsilon
_{d}=0.80$~MeV, $A = -0.025$~MeV,  $ B = -0.01$~MeV and $ C =
0.004$~MeV, with Eq.~(7) we can fix the critical angular momentum
$L_c=8 \hbar $. The energy of the yrast states against the angular
momentum $L$ can be illustrated in the left panel of Fig.~3. It
has been known that the energy of E2 transition $\gamma$-ray over
spin (E-GOS)  $ R = \frac{E_{\gamma}(L \rightarrow L-2)}{L} $ can
be taken as a quite good signature to manifest the vibrational to
axially rotational phase transition along the yrast
line\cite{Regan03}. As an auxiliary evidence, we show also the
E-GOS of the yrast states with these parameters in the right panel
of Fig.~3. The figure indicates apparently that the yrast states
involve a vibrational to axially rotational phase transition and
the angular momentum $L_c$ is definitely the critical point for
the phase transition to take place. Such a transition is quite
similar to that between the states with $n_p=0$ and
$n_p=N$\cite{BF02} in the vibron model\cite{I78,IL95} with random
interactions.

In the case $A<0$, $B>0$ or $A>0$, $B<0$, there may exist a
critical angular momentum $L_c$ with $ \frac{L_c}{2} \le N_0 < N$,
too. For the states with $L \le L_c$, if $\varepsilon_{d} + A
(N_{0} +4) + B(N_0 +3) > 0 $, their energies behave like an
anharmonic vibrator's. For the ones with $L > L_c$, their energies
behave as the quasi-rotational states with $N_0$ bosons. It is
also possible that the $N_0$ takes several values less than the
$N$. Then the second and more higher backbending may emerge.

We have analyzed the available experimental data of the yrast
states of the nuclei with $ 30\le Z \le 100$ and simulated the
data with least-square fitting in our present approach. Both the
experimental
data\cite{PJJS01,FJ98,Blac00,FJ96,Blac01,Singh011,Singh012,Singh022,Zhou98}
and the best fitted energy spectra of the yrast states in some
even-even nuclei are illustrated in Fig.~4 (the other data are
available if required). It shows obviously that, besides the ones
identified in the $A \sim 110$ mass region\cite{Regan03}, there
exist vibrational to axially rotational phase transition along the
yrast line in other nuclei. For example, $^{72}$Se, $^{90}$Zr,
$^{92}$Zr, $^{98}$Pd, $^{108}$Sn, $^{116}$Sn, $^{136}$Te,
$^{146}$Gd, $^{148}$Gd, $^{148}$Dy, $^{150}$Dy, $^{186}$Hg,
$^{194}$Pb, $^{196}$Pb, $^{198}$Pb, $^{200}$Po, $^{202}$Po,
$^{206}$Po, $^{214}$Rn and others are the ones whose anharmonic
vibrational frequency in the corresponding phase decreases against
the increasing of angular momentum. While $^{66}$Zn, $^{66}$Ge,
$^{68}$Se, $^{82}$Sr, $^{84}$Sr, $^{86}$Mo, $^{114}$Pd,
$^{116}$Pd, $^{116}$Cd, $^{118}$Cd, $^{120}$Cd, $^{112}$Te,
$^{114}$Te, $^{116}$Te, $^{116}$Xe, $^{118}$Xe, $^{124}$Xe,
$^{128}$Xe, $^{130}$Xe, $^{124}$Ba, $^{130}$Ba, $^{148}$Sm,
$^{142}$Gd, $^{144}$Dy, $^{156}$Er, $^{158}$Yb, $^{160}$Yb,
$^{160}$Hf, $^{162}$Hf, $^{164}$W, $^{170}$Os, $^{188}$Pt,
$^{192}$Pt, $^{194}$Pt, $^{184}$Hg and so on are the ones whose
vibrational frequency increases against the increasing of angular
momentum.

In summary, by analyzing the energy surface of the nucleus in U(5)
symmetry in the framework of thermodynamics, we give a phase
diagram of the vibration and the rotation in terms of the angular
momentum and the deformation parameter in this letter. We have
then proposed an approach to describe the vibrational to axially
rotational phase transition along the yrast line. With the energy
spectrum being examined, we show that our presently proposed
approach can describe very well the vibrational to axially
rotational phase transition along the yrast line.  We have also
analyzed the energy spectra of the even-even nuclei with $ 30 \le
Z \le 100$. It shows that, besides the ones in $A \sim 110$ mass
region identified by Regan and collaborators\cite{Regan03}, there
exist many other nuclei involving vibrational to axially
rotational phase transition along the yrast line.

\vspace*{8mm}

This work is supported by the National Natural Science Foundation
of China under the contract Nos. 19875001, 10075002, and 10135030
and the Major State Basic Research Development Program under
contract No. G2000077400. One of the authors (Y.X. Liu) thanks the
support by the Foundation for University Key Teacher by the
Ministry of Education, China, too.


\newpage

\parindent=0pt


\begin{figure}
\begin{center}
\includegraphics[scale=0.50,angle=-90]{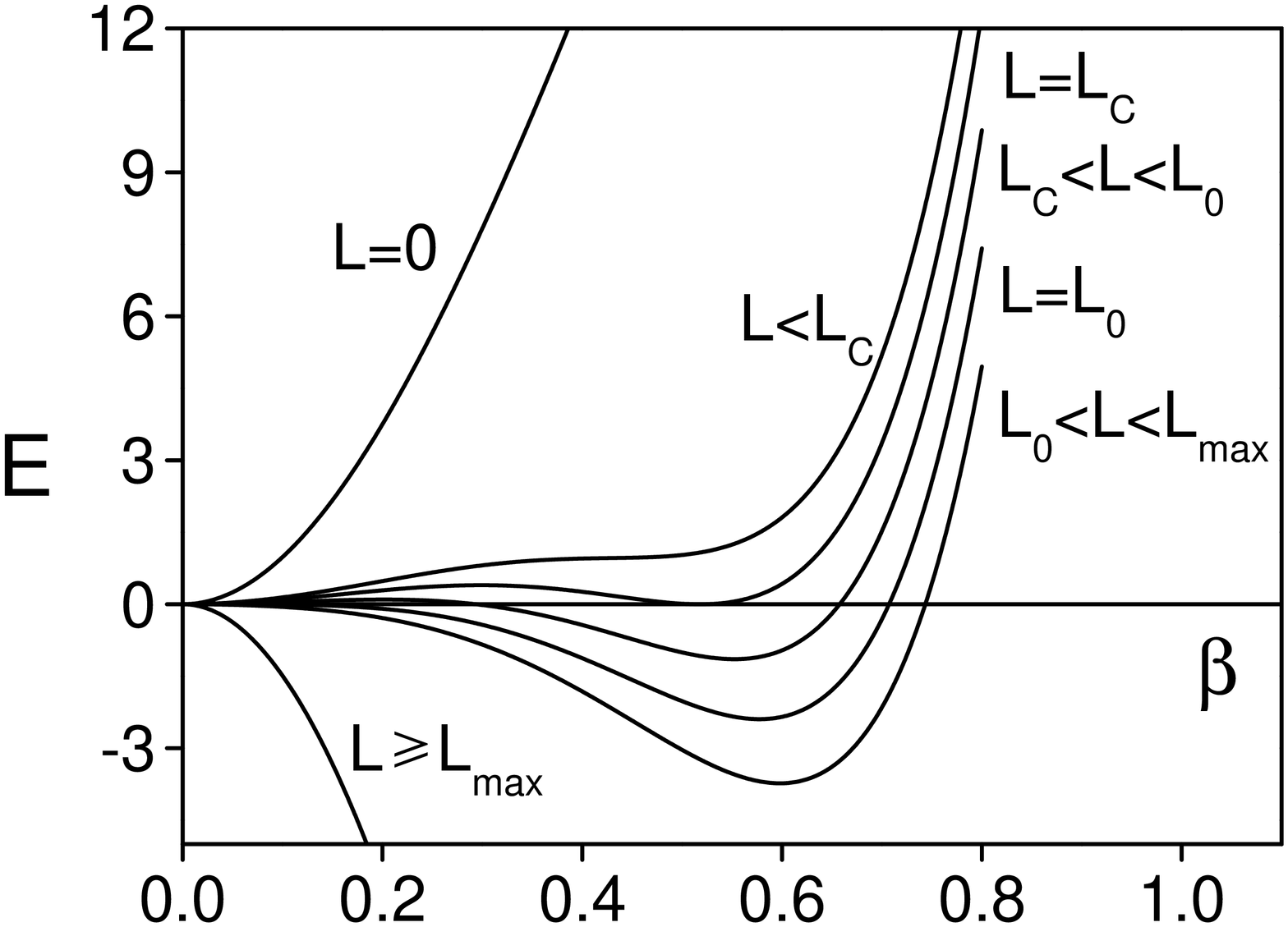}
\caption{The energy surface (Landau free energy) of a nucleus
against the ``deformation parameter" $\beta$ at some typical
angular momentum $L$ (with the parameters in Eq.~(5) being taken
as $E_0 =0$, $L_0 = 20$, $\alpha = -10.0000$, $AN(N-1) =
-64.6667$, (in arbitrary unit) ). }
\end{center}
\end{figure}

\begin{figure}
\begin{center}
\includegraphics[scale=0.50,angle=-90]{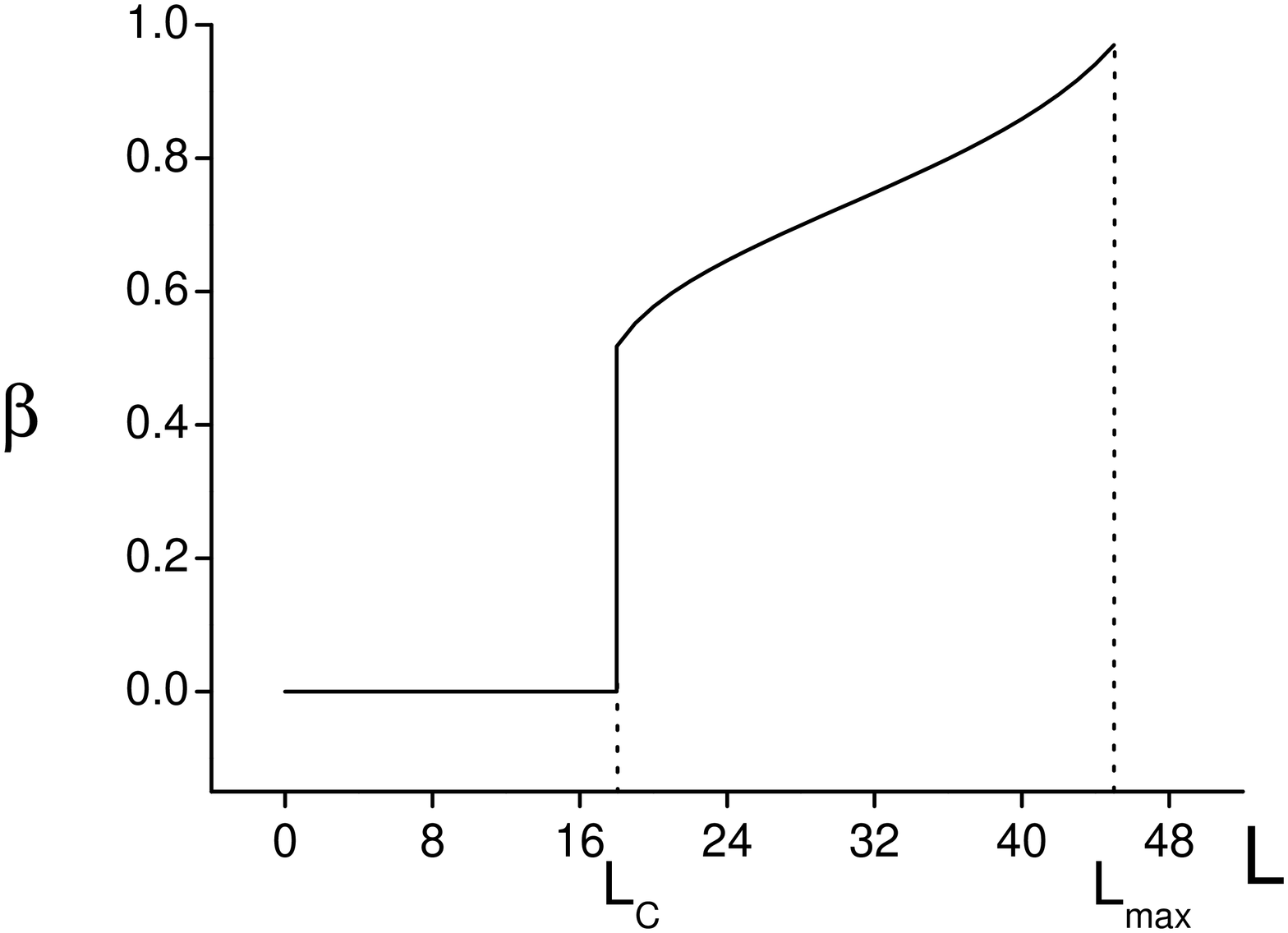}
\caption{The vibration and rotation phase diagram of a nucleus in
terms of the angular momentum $L$ and the ``deformation parameter"
$\beta$ (with the same parameters for Fig.~1). }
\end{center}
\end{figure}

\begin{figure}
\begin{center}
\includegraphics[scale=0.50,angle=-90]{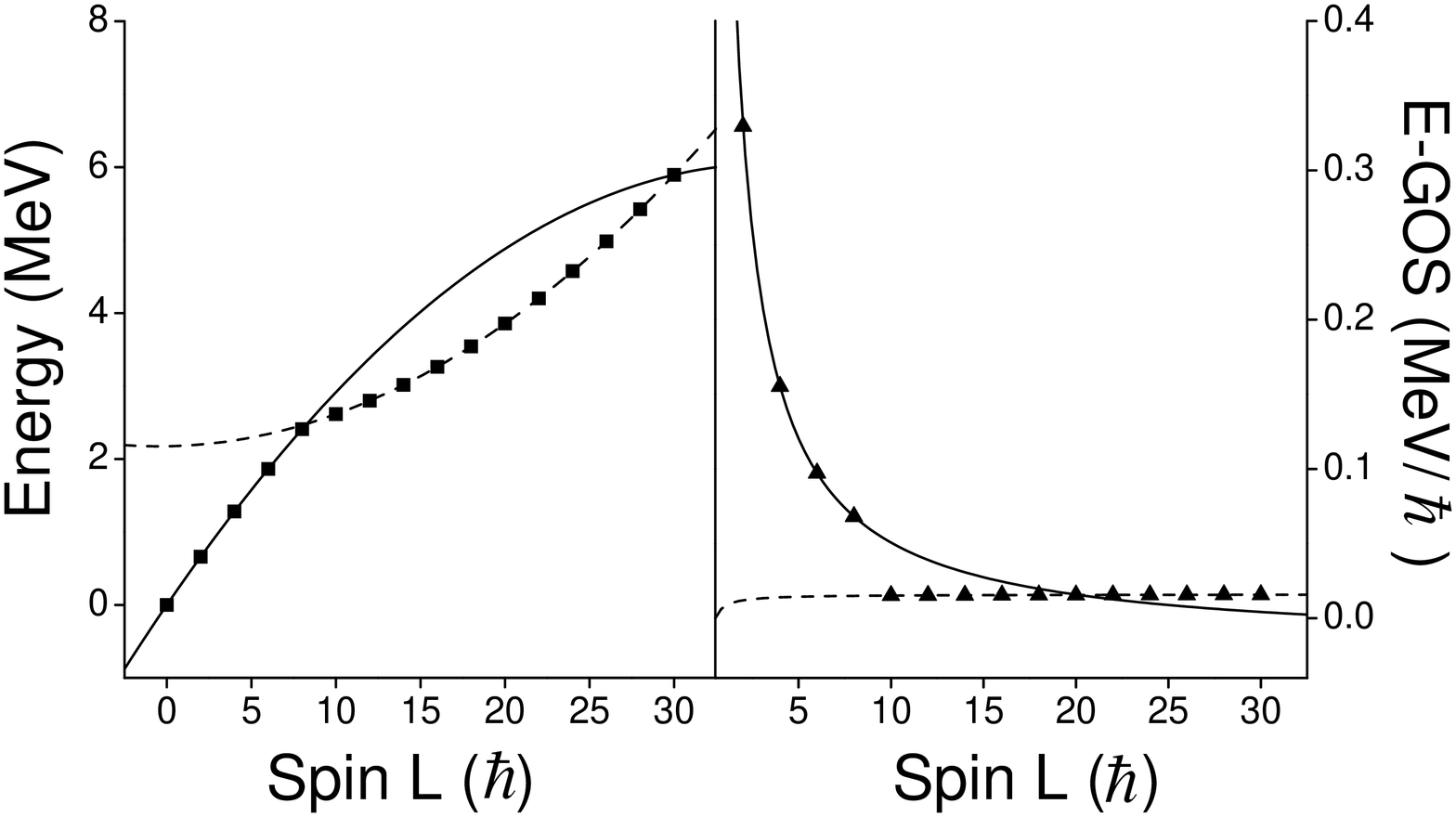}
\caption{An example of the energy against spin (left panel) and
the E-GOS (right panel) along the yrast line (filled diamonds and
triangles, respectively) in the approach of U(5) symmetry with $A
< 0$ (with parameters $\varepsilon _d = 0.80$~MeV, $A= -
0.025$~MeV, $B = -0.01$~MeV, $C= 0.004$~MeV. The solid and dashed
lines are implemented to guide the eye.)}
\end{center}
\end{figure}

\vfill

\begin{figure}
\begin{center}
\includegraphics[scale=0.59,angle=0]{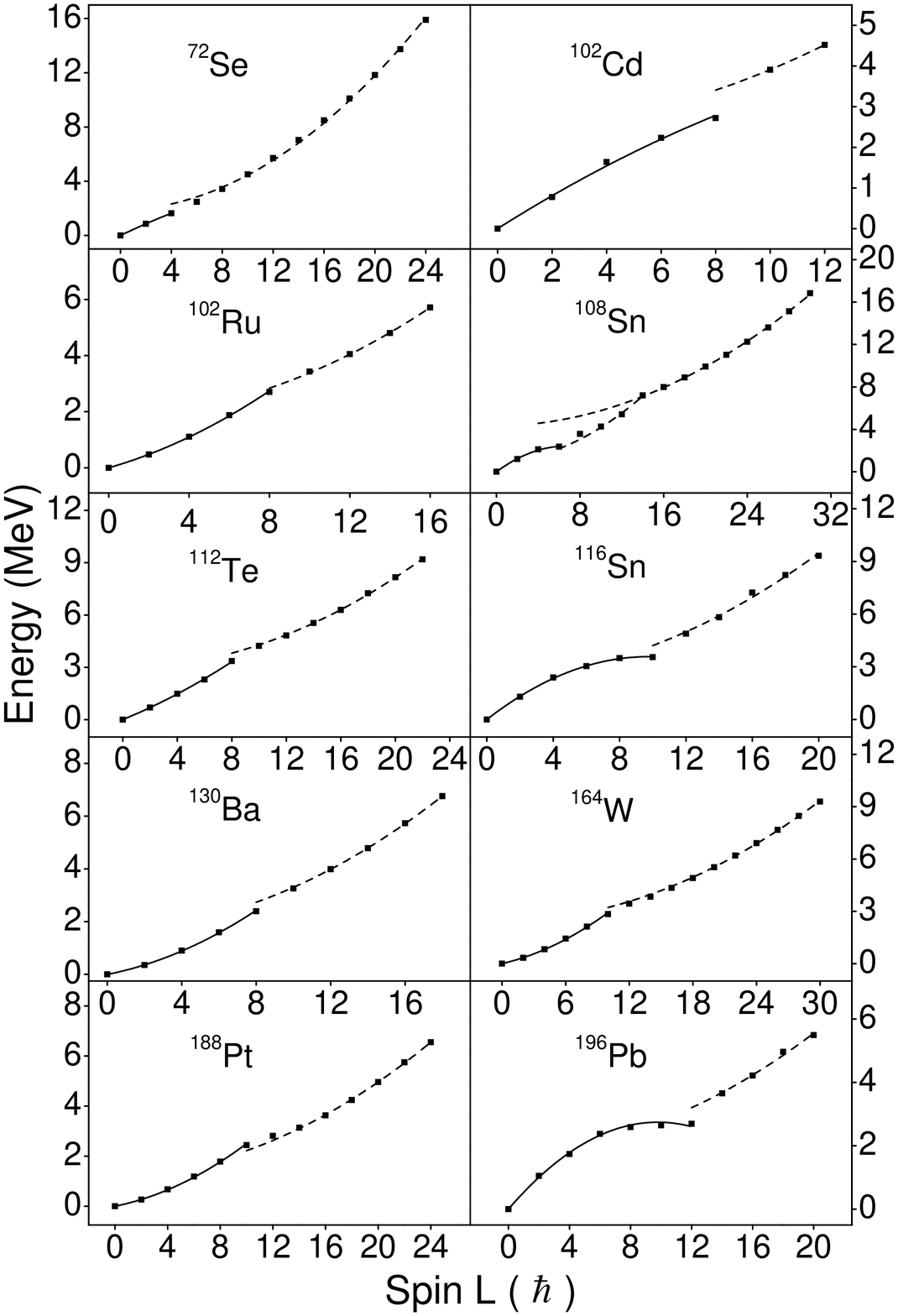}
\caption{The experimental data of the energy against spin (filled
squares) of some even-even nuclei and the simulated results in the
present approach (solid, dashed curves for vibrational, rotational
mode, respectively).}
\end{center}
\end{figure}

\end{document}